\documentclass[aps,prb,twocolumn,floatfix,superscriptaddress,longbibliography]{revtex4-1}
\usepackage{graphicx,txfonts,bm}
\usepackage[colorlinks,citecolor=magenta,linkcolor=blue]{hyperref}
\usepackage{hyperref}
\usepackage{natbib}

\begin{document}

\title{Correlated 5$f$ electronic states and phase stability in americium under high pressure: Insights from DFT+DMFT calculations}
\author{Haiyan Lu}
\email{hyluphys@163.com}
\affiliation{Institute of Materials, China Academy of Engineering Physics, Huafengxincun No.~9, Mianyang 621907, Sichuan, China}
\date{\today}

\begin{abstract}
We investigate the electronic structure of americium (Am) across its four experimentally confirmed high-pressure phases Am-I ($P6_3/mmc$), Am-II ($Fm\bar{3}m$), Am-III ($Fddd$), and Am-IV ($Pnma$) up to 100 GPa, using density functional theory combined with embedded dynamical mean-field theory. Our results successfully reproduce the prominent localized 5$f$ peak observed in ultraviolet photoelectron spectroscopy around –2.8~eV below the Fermi level in the Am-I phase. While 5$f$ electrons in Am-I and Am-II remain strongly localized, those in Am-III and Am-IV manifest discernible signatures of increased hybridization: a noticeable shift of spectral weight toward the Fermi level, enhanced hybridization strength, and the emergence of distinct multi-peak structures. These changes indicate that 5$f$ electrons begin to participate in bonding and undergo partial delocalization under pressure. Nevertheless, the spectral weight of 5$f$ electrons near the Fermi level in Am-IV remains relatively low, indicating that, compared to U and Pu, Am retains stronger localized 5$f$ electrons even under high pressure. Analysis of the electronic configurations reveals pressure-enhanced valence state fluctuation, characterized by the mixing of $5f^{5}$, $5f^{6}$, and $5f^{7}$ electronic configurations. The X-ray absorption branching ratio further shows that the angular-momentum coupling scheme approaches the $jj$ limit. Additionally, we demonstrate that the stability of the low-symmetry high-pressure phases (Am-III and Am-IV) is governed by a Peierls-like distortion mechanism, which reduces the total energy through symmetry-lowering lattice distortions accompanied by electronic reconstruction. This study offers a new microscopic perspective on high-pressure phase transitions and emergent quantum phenomena in actinide materials.
\end{abstract}

\maketitle
\section{Introduction\label{sec:introduction}}

In actinides, the partially filled 5$f$ subshell gives rise to a rich spectrum of emergent quantum phenomena~\cite{shim:2007,nature4202002,Science1978AmSC}, strikingly manifested in a pronounced volume jump that unfolds progressively across from plutonium (Pu) to americium (Am), as the 5$f$ electron count increases~\cite{RevModPhys.81.235}. This discontinuous volumetric expansion directly signals a marked enhancement in the localization of 5$f$ electrons in Am. Indeed, Am represents the first transuranic actinide in which 5$f$ electrons form a closed relativistic subshell. Governed by strong spin–orbit coupling, the 5$f$ orbitals split into $j=5/2$ and $j=7/2$ manifolds. The six 5$f$ electrons predominantly occupy the $j=5/2$ subshell, resulting in a non-magnetic, atomic-like 5$f^{6}$ configuration. Thich produces a $^{7}F_0$ singlet ground state with $jj$ angular momentum coupling scheme, characterized by total angular momentum $J=0$~\cite{PhysRevLett.94.097002,PhysRevB.76.073105}.

Pressure serves as a crucial experimental parameter for tuning the localization degree of 5$f$ electrons and driving structural phase transitions in Am. By increasing the overlap between 5$f$ orbitals, pressure may promote delocalization of 5$f$ states. At ambient pressure, Am crystallizes in a double hexagonal close-packed structure (space group $P6_3/mmc$), designated as Am-I. Upon compression, it first transforms to a face-centered cubic phase ($Fm\bar{3}m$, Am-II) above 6.1~GPa, then into a face-centered orthorhombic structure ($Fddd$, Am-III) at 10.0~GPa, followed by a subsequent transformation to a primitive orthorhombic phase ($Pnma$, Am-IV) around 17.6~GPa. Especially, the last two phases exhibit low symmetry and relatively reduced volumes, a signature of enhancing 5$f$ electron participation in bonding and manifestation of pressure-induced incremental delocalization~\cite{heathman2000,PhysRevB.63.214101}.

The transport properties of Am, including electrical resistivity and specific heat~\cite{J.Low.Temp.Phys.30.561,LINK1994148,STEPHENS1968815,SCHENKEL19771301}, directly reflect the pressure-dependent behavior of 5$f$ electrons. Measurements reveal a pronounced increase in resistivity with both rising pressure~\cite{LINK1994148,STEPHENS1968815} and temperature~\cite{LINK1994148,SCHENKEL19771301}, a hallmark of 5$f$ localization that distinguishes Am from lighter actinides. Under compression, however, these electrons gradually delocalize: resistivity studies indicate a shift from a 5$f^{6}$ toward 5$f^{7}$ electronic configuration, signaling pressure-driven valence state fluctuations~\cite{EPL82.57007}. This delocalization is further mirrored in the superconductivity in Am, which emerges near 0.8~K at ambient pressure~\cite{Science1978AmSC}, reaches a maximum $T_c$ of 2.3~K around 10~GPa, and then declines upon further compression~\cite{LINK1994148}, closely tracking the sequence of structural phase transitions. The evolution of both normal-state resistivity and superconductivity can be attributed to the progressive involvement of 5$f$ electrons in bonding under pressure~\cite{GRIVEAU200784}. Accordingly, the 5$f$ states play a decisive role in governing the transport behavior, unveiling the localized 5$f$ electrons strongly scattering the $spd$ conduction electrons, whereas the itinerant 5$f$ electrons contributing to conduction. Consequently, Am provides a quintessential platform for understanding actinides, exemplifying how pressure drives the 5$f$ electrons from a localized, atomic-like state into a delocalized, bonding-active state.

The strong localization of 5$f$ electrons in Am-I, evidenced by a pronounced spectral feature around –2.8~eV, is well established through X-ray and ultraviolet photoemission spectroscopy at ambient pressure~\cite{PhysRevLett.52.1834,PhysRevB.45.13239,PhysRevB.72.115122}. Nevertheless, whether and when these electrons become itinerant under high pressure remains contentious. Resonant X-ray emission spectroscopy and X-ray absorption near‑edge structure measurements at the Am $L_3$ edge up to 23~GPa reveal no detectable mixed‑valence behavior or valence transition between 5$f^{6}$ and 5$f^{7}$ configurations~\cite{PhysRevB.82.201103}. This stands in contrast to high-pressure structural and transport studies~\cite{heathman2000,PhysRevB.63.214101,EPL82.57007}, as well as certain theoretical predictions~\cite{JPCM17.257,PhysRevB.72.024109}, which suggest that pressure should induce valence state fluctuations and a progressive delocalization of 5$f$ electrons. 
Furthermore, estimates of the volume collapse expected from 5$f$ delocalization vary widely from about 34 \%~\cite{PhysRevB.45.3198} to 25 \%~\cite{PhysRevB.61.8119}, reflecting persistent uncertainty in the predicted electronic response. These discrepancies highlight two unresolved issues: first, whether 5$f$ electrons in Am become genuinely itinerant under compression; and second, if so, at which structural phase (Am-II, Am-III, or Am-IV) this itinerancy begins. Clarifying this pressure-induced localized to itinerant crossover calls for further integrated experimental and theoretical investigation.

To elucidate the pressure-driven evolution of 5$f$ electrons in americium, a reliable theoretical framework capable of capturing strong electron correlations is essential. Most previous theoretical studies, however, have focused on idealized high-symmetry structures (typically fcc) and often neglected spin-orbit coupling. Conventional density functional theory (DFT) and its DFT+$U$ extensions which are constrained by a single-particle picture, fail to reproduce the experimental photoemission spectrum of Am-I~\cite{PhysRevB.84.075138,PhysRevB.73.104415,J.AlloysCompd.444.42,SolidStateCommun.150.938,SolidStateCommun.164.22} and frequently predict magnetic ground states for Am-I and Am-II, contradicting the observed non‑magnetic behavior~\cite{JPCM14.3575,JPCM17.257,PhysRevB.72.024109,PhysRevB.45.3198}. Although hybrid density functionals~\cite{Chem.Phys.Lett.482.223} improve the agreement with photoemission data~\cite{PhysRevB.72.115122}, they still do not fully account for the many-body correlations inherent to 5$f$ electrons.
A more advanced approach, DFT combined with dynamical mean-field theory (DFT + DMFT) has proven successful in describing strongly correlated materials and reproducing key spectroscopic features~\cite{PhysRevB.101.125123,PhysRevB.103.205134,PhysRevB.109.205132}. Early DFT + DMFT implementations for Am, however, relied on several simplifications: the total energy was not computed self-consistently~\cite{PhysRevLett.96.036404,PhysRevB.94.115148}, the quantum impurity problem was treated within the one-crossing approximation (OCA)~\cite{PhysRevB.64.155111}, and calculations were typically performed on an idealized fcc lattice rather than on experimentally determined crystal structures. Consequently, electronic properties such as the density of states, band dispersion, and Fermi surface derived from these studies remain incomplete and do not yet offer a realistic, pressure‑resolved picture of Am's electronic behavior.

In this study, we comprehensively investigate the electronic structures of the four experimentally established phases of americium (Am-I, Am-II, Am-III, and Am-IV) using a charge fully self-consistent DFT + DMFT approach~\cite{RevModPhys.68.13,RevModPhys.78.865,PhysRevB.81.195107} implemented with the continuous-time hybridization expansion quantum Monte Carlo (CT-HYB) impurity solver. Section~\ref{sec:method} outlines the computational framework of the DFT + DMFT methodology. In Section~\ref{sec:results}, we first validate our approach by reproducing the experimental density of states, and then elucidate the pressure-driven evolution of 5$f$ electrons from localized enhancing itinerant character, including an analysis of electronic configuration distributions and valence state fluctuation behavior. We also examine the strength of 5$f$ electronic correlations and the evolution of the angular momentum coupling scheme. Section~\ref{sec:discussion} examines the role of Peierls-type lattice distortion in stabilizing the low-symmetry high-pressure phases. Finally, a concise summary is presented in Section~\ref{sec:summary}.

\section{Methods\label{sec:method}}
In order to capture the correlation between 5$f$ electrons of americium, a combination of the density functional theory and the embedded dynamical mean-field theory (DFT + DMFT) is employed. This approach integrates the realistic band structure calculation from DFT and a non-perturbative treatment of the many-body local interaction effects within DMFT~\cite{RevModPhys.68.13,RevModPhys.78.865}.

\textbf{DFT calculations.} We employed the \texttt{WIEN2K} code to perform DFT calculations, which implements a full-potential linearized augmented plane-wave formalism~\cite{wien2k}. The experimental crystal structure of Am was adopted~\cite{heathman2000}. The muffin-tin sphere radii for Am atom was chosen 2.50 au, with $R_{\text{MT}}K_{\text{MAX}} = 8.0$. The valence states included the $5f$, $6d$, and $7s$ orbitals for Am. The exchange-correlation potential was evaluated using the Generalized Gradient Approximation (GGA) with the Perdew-Burke-Ernzerhof (PBE) functional~\cite{PhysRevLett.77.3865}. The spin-orbit coupling effect was included in a variational manner. Am was treated as nonmagnetic according to the results of high pressure experiment~\cite{heathman2000}.

\textbf{DFT + DMFT calculations.} We utilized the \texttt{eDMFT} software package to perform the DFT + DMFT calculations~\cite{PhysRevB.81.195107}. The calculated temperature is $T\approx290$~K, where the inverse temperature $\beta$=40 is defined in units of $(k_B\cdot T)^{-1}$, with $k_B$ being the Boltzman constant and $T$ the temperature in Kelvin. The correlated nature of the Am-$5f$ orbitals was treated within the DMFT formalism. The Coulomb interaction matrix for Am-$5f$ orbitals was constructed using Slater integrals, with the Coulomb repulsion interaction parameter $U$=5.0~eV and Hund's exchange interaction parameter $J_{\text{H}}$=0.6~eV. The local impurity Hamiltonian was formulated in the $|J, J_z\rangle$ basis. The DMFT projectors, which map the Kohn-Sham states onto the localized impurity basis, were constructed within a large energy window spanning from -10~eV to 10~eV relative to the Fermi level.
The constructed multi-orbital Anderson impurity models were solved using the hybridization expansion continuous-time quantum Monte Carlo impurity solver (dubbed as CT-HYB)~\cite{RevModPhys.83.349,PhysRevLett.97.076405,PhysRevB.75.155113}. 
To maintain computational tractability, the local Hilbert space was truncated by retaining only atomic eigenstates with electron occupancies $N \in [5,7]$. The double-counting correction is essential to avoid counting electron-electron interactions both in the DFT exchange-correlation functional and explicitly added in the Hubbard term. Here we chose the fully localized limit (FLL) scheme~\cite{jpcm:1997}, which is derived from the atomic limit and is widely recognized as appropriate for strongly correlated electron systems. The fully localized limit double-counting term is given by 
\begin{equation}
\Sigma_{dc} = U \left(n_{5f} - \frac{1}{2}\right) - \frac{ J_{H} } {2} \left(n_{5f} -1 \right),
\end{equation}
where $n_{5f}=6.0$ is chosen as the nominal occupancy of Am-$5f$ orbitals during the DFT + DMFT calculations, consistent with theoretical studies~\cite{PhysRevLett.96.036404}. Charge fully self-consistent DFT + DMFT calculations were performed, requiring approximately $60 \sim 80$ DFT + DMFT iterations to achieve good convergence. The convergence criteria for charge density and total energy were set to $10^{-4}$~e and $10^{-4}$~Ry, respectively. The final output were Matsubara self-energy function $\Sigma(i\omega_n)$ and impurity Green's function $G(i\omega_n)$, which were then utilized to obtain the integral spectral functions $A(\omega)$ and momentum-resolved spectral functions $A(\mathbf{k},\omega)$. From the probability of atomic eigenstates, we can extract key information concerning the 5$f$ valence state fluctuation and electronic configuration.

\section{Results\label{sec:results}}
\subsection{Comparison with experimental photoemission spectroscopy} 

\begin{figure}[ht]
\includegraphics[width=0.5\textwidth]{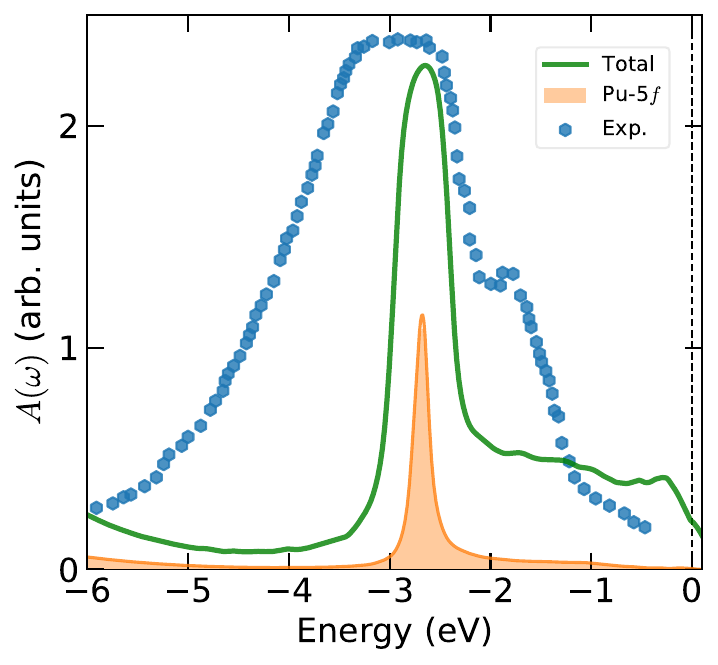}
\caption{(Color online) Comparisons of theoretical and experimental density of states for Am. The total density of states (green-solid line) after considering the Fermi-Dirac distribution and 5$f$ partial density of states (color-filled regions) of Am-I phase at $T = 290$~K obtained by the DFT + DMFT method. The experimental data (filled blue circles) are taken from ~\citet{PhysRevLett.52.1834}. The Fermi level $E_F$ is represented by vertical dashed line. \label{fig:doscom}} 
\end{figure}

To validate the computational methodology, Fig.~\ref{fig:doscom} presents the calculated total electronic density of states of the Am-I phase and experimental ultraviolet photoelectron spectroscopy~\cite{PhysRevLett.52.1834}. For a direct comparison with the experimental measurements performed at 290~K, the computed total density of states was multiplied by the Fermi-Dirac distribution function $f(\epsilon)=1/[exp(\epsilon\beta)+1]$ with the inverse temperature $\beta$=40 (corresponding to $T\approx290$~K). The resulting Am-5$f$ electron density of states, weighted by the Fermi-Dirac distribution at 290~K, exhibits a prominent peak at approximately –2.8~eV below the Fermi level, indicating strongly localized 5$f$ electronic states. The consistency between the computed and experimental spectra confirms the reliability of our computational approach. 

\subsection{Quasiparticle resonance peaks}

\begin{figure*}[ht]
\includegraphics[width=\textwidth]{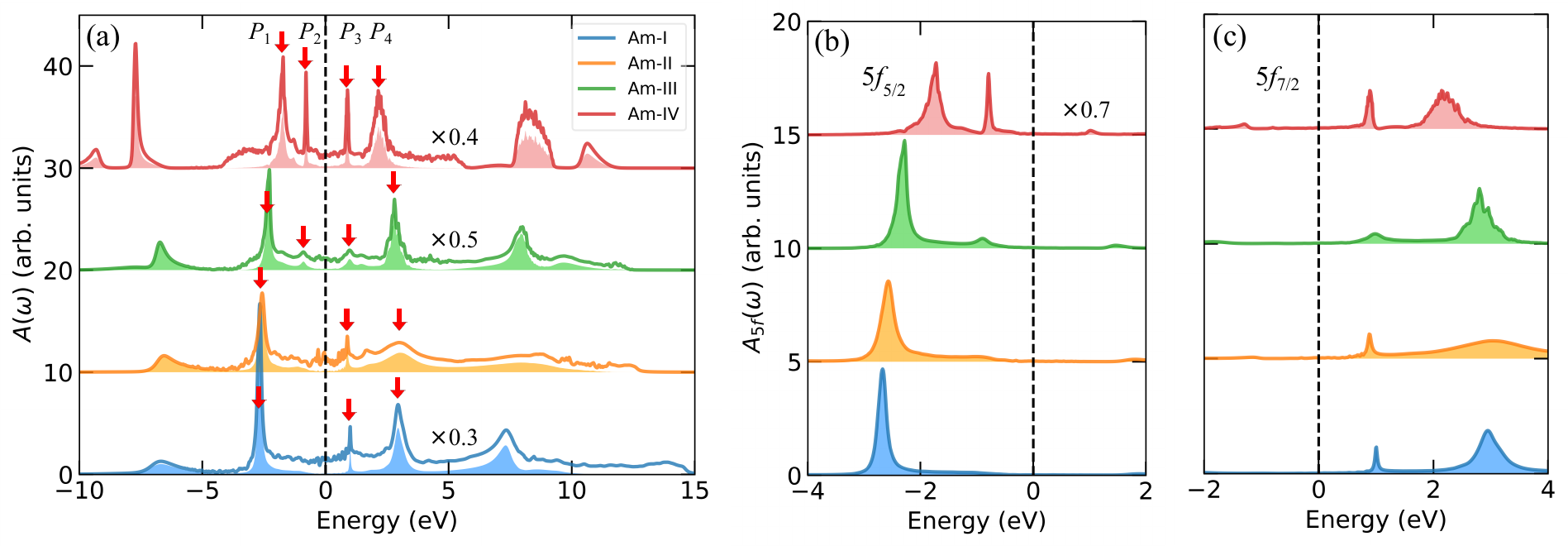}
\caption{(Color online) Total and 5$f$ partial density of states of Am at $T = 290$~K by DFT + DMFT calculations. (a) Total density of states A($\omega$) (in solid lines) and 5$f$ partial density of states A$_{5f}$($\omega$) (in colored shadow regions). (b) and (c) Orbital-resolved (or $j$-resolved) 5$f$ partial density of states $j$=5/2 and $j$=7/2. The Fermi levels $E_{F}$ are represented by vertical dashed lines. \label{fig:dos}}
\end{figure*}

The evolution of the electronic density of states across the four phases of Am-I, Am-II, Am-III, and Am-IV [Fig.~\ref{fig:dos} (a)] reveals a systematic pressure-induced delocalization of the 5$f$ electrons and their enhanced hybridization with conduction electrons. In the ambient-pressure Am-I phase, the 5$f$ states are highly localized, manifested as a sharp peak ($P_1$) centered near –2.8~eV. As illustrated in Fig.~\ref{fig:dos} (b) and (c), the $P_1$ peak is predominantly contributed by the occupied $j$ = 5/2 states, corresponding to the transition 5$f^{6} (^{7}F_0)\rightarrow$ 5$f^{5} (^{6}H_{5/2})$. The density of states near the Fermi level is nearly negligible, consistent with a non‑magnetic ground state of total angular momentum $J$ = 0. Notably, two weaker peaks ($P_3$ and $P_4$) observed above the Fermi level at approximately 1.0~eV and 3.0~eV. These features originate mainly from unoccupied $j$ = 7/2 states, associated with the transitions 5$f^{6} (^{7}F_0)\rightarrow$ 5$f^{7} (^{8}S_{7/2})$ and 5$f^{6} (^{7}F_0)\rightarrow$ 5$f^{7} (^{6}P_{7/2})$, respectively.

Upon transitioning to the Am-II phase, the main peak $P_1$ shifts slightly toward the Fermi level to –2.6~eV, accompanied by a gradual transfer of spectral weight to energies closer to the Fermi level. The 5$f$ electrons remain predominantly localized. Similar to the Am-I phase, $P_3$ and $P_4$ peaks are observed above the Fermi level at about 0.9~eV and 3.0~eV, again attributed to unoccupied $j$ = 7/2 states, with the $P_4$ peak exhibiting a broader spectral envelope.
In the Am-III phase, a distinct shoulder peak ($P_2$) emerges at –0.9~eV adjacent to the main peak $P_1$ centered at –2.3~eV. Concurrently, multiple resonant features ($P_3$ and $P_4$) develop above the Fermi level at 1.0~eV and 2.8~eV, demonstrating atomic-multiplet characteristics reminiscent of those observed in Pu. This marks the onset of 5$f$ electron delocalization and their participation in bonding.

At even higher pressure in the Am-IV phase, the main peak $P_1$ moves further upward to -1.8~eV, while the intensity of the $P_2$ peak at -0.8~eV increases substantially. Above the Fermi level, the $P_3$ peak at 0.9~eV gains significant intensity, and the $P_4$ peak shifts closer to the Fermi level, settling at 2.1~eV. This leads to split double-peak structures for both the $j$ = 5/2 and $j$ = 7/2 manifolds. Additionally, lower and upper Hubbard bands are identified in the energy ranges (–10~eV, –5~eV) and (5~eV, 15~eV), respectively. These spectral changes signal the onset of 5$f$ electron itinerancy and hybridization with conduction electrons ($f-c$). Starting from the Am-III phase, a systematic transfer of spectral weight and the emergence of multiple peak structures reflect the gradual delocalization of 5$f$ electrons in the low-symmetry phases stabilized by Peierls distortions. This evolution marked by the increasing participation of 5$f$ electrons in bonding confirms that from Am-III onward, the 5$f$ states become progressively less localized and engage in chemical bonding~\cite{GRIVEAU200784}. Even in the Am-IV phase, however, the 5$f$ electrons retain a considerable degree of localization. The trend is further corroborated by the observed rise in electrical resistivity with pressure~\cite{PhysRevLett.94.097002}, a consequence of enhanced scattering between conduction electrons and the increasingly delocalized 5$f$ states.

\subsection{Band structure and hybridization function} 

\begin{figure*}[ht]
\includegraphics[width=\textwidth]{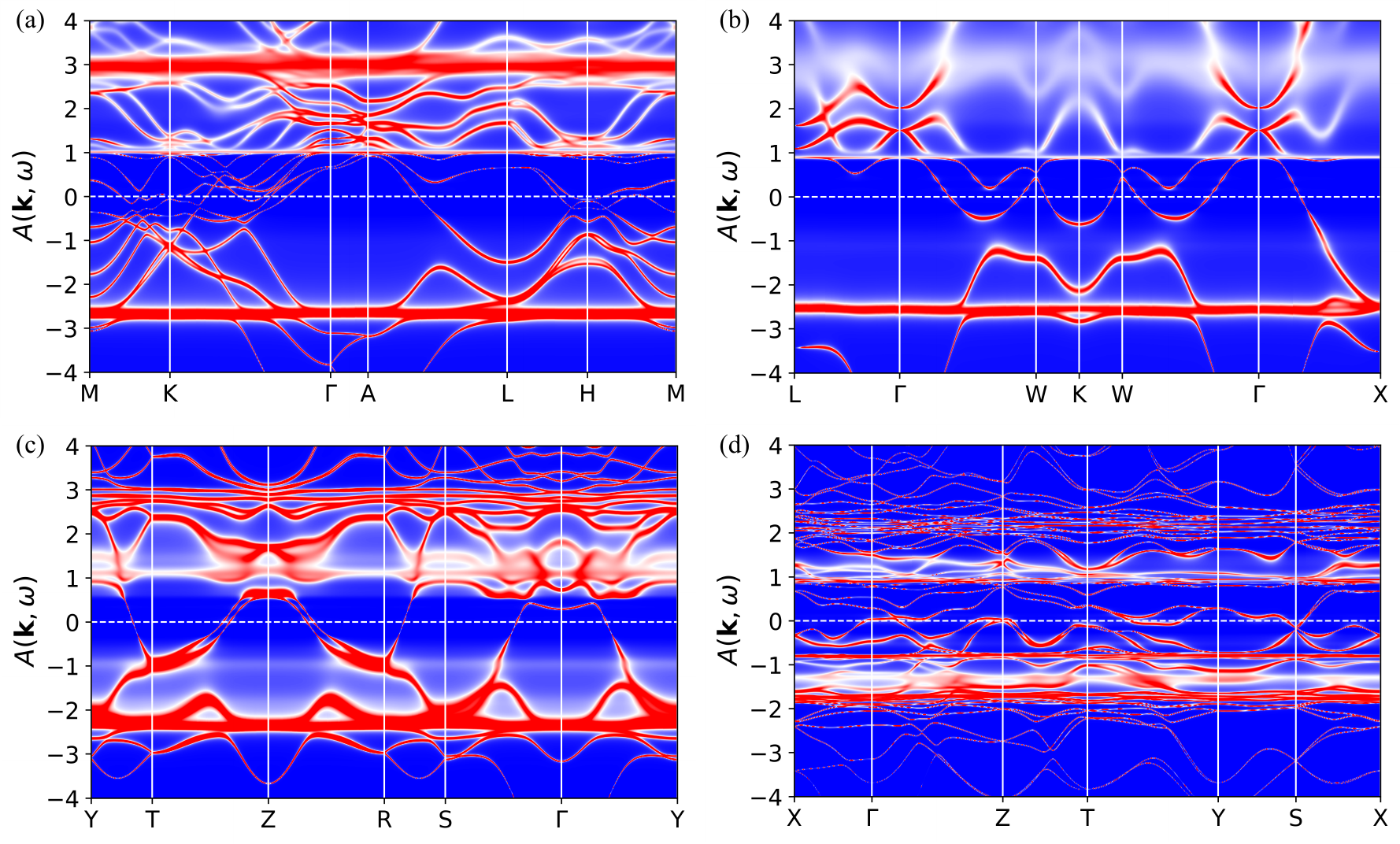}
\caption{(Color online). Momentum-resolved spectral functions A(${\bf k}$, $\omega$) of Am at $T = 290$~K obtained by the DFT + DMFT method. (a) Am-I. (b) Am-II. (c) Am-III. (d) Am-IV. The horizontal dashed lines denote the Fermi levels.
\label{fig:akw}}
\end{figure*}

\begin{figure*}[ht]
\includegraphics[width=\textwidth]{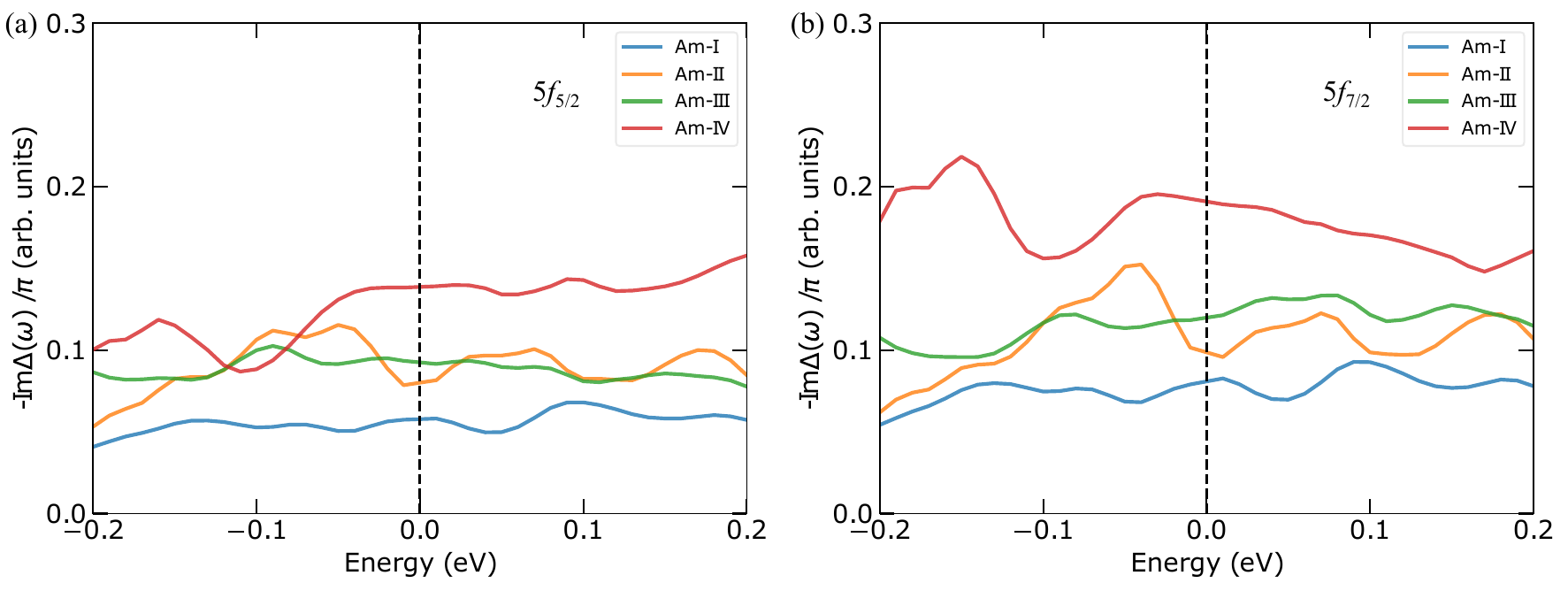}
\caption{(Color online) Hybridization functions of Am at $T = 290$~K obtained by the DFT + DMFT method. (a) For the Am-5$f$ $j$=5/2 components. (b) For the Am-5$f$ $j$=7/2 components. \label{fig:hyb}}
\end{figure*}

The evolution of the electronic band structure along high-symmetry paths in the Brillouin zone corroborates the density of states calculations. Figures~\ref{fig:akw}(a)–(d) display the electronic bands along high-symmetry $k$-points for Am-I, Am-II, Am-III, and Am-IV, respectively. In the Am-I phase, the spectral weight of Am-5$f$ electrons near the Fermi level is nearly negligible; it is instead distributed mainly at about –3~eV below and around 1~eV and 3~eV above the Fermi level. Owing to $f-c$ hybridization, the lower and upper Hubbard bands are broadened into relatively flat ``hump'', exhibiting characteristics of many-body states. Although the previous DFT + DMFT calculation~\cite{PhysRevLett.96.036404} employed a different fcc lattice and high-symmetry $k$-path for Am, leading to differences in band shapes, the energy regions in which the 5$f$ bands locate in our calculation agree with those results~\cite{PhysRevLett.96.036404}. This confirms that at ambient pressure the 5$f$ electrons remain strongly localized, with spectral weight far from the Fermi level. 
As pressure increases, the system stabilizes in the Am-II lattice. Our calculated band shapes are largely consistent with the earlier study~\cite{PhysRevLett.96.036404}, showing that the 5$f$ bands are concentrated around –2.8~eV below the Fermi level and near 1.0~eV and 3.0~eV above it. At this stage, the 5$f$ electrons still reside well away from the Fermi level, retaining strong localization.

In the Am-III phase, the spectral weight of the 5$f$ electrons shifts noticeably toward the Fermi level, with a pronounced distribution at –2.3~eV and an emerging feature at –0.9~eV. Moreover, flat bands appear at about 1.0~eV and 2.8~eV above the Fermi level, displaying atomic-multiplet-like characteristics similar to those seen in plutonium. These changes mark the onset of 5$f$ electron delocalization and their participation in bonding. 
In the higher-pressure Am-IV phase, the number of flat 5$f$ bands increases further, and the spectral weight continues to approach the Fermi level. Prominent spectral weights appear at –1.8~eV, -0.8~eV, 0.9~eV, and 2.1~eV, accompanied by a lower Hubbard band in the range (–10~eV, –5~eV) and an upper Hubbard band in (5~eV, 15~eV). These features indicate the enhancing itinerant 5$f$ electrons and hybridization with conduction electrons. Together with the density of states, the transfer of spectral weight and the formation of multiple-peak structures from Am-III to Am-IV reflect a gradual delocalization of the 5$f$ electrons, which is stabilized by Peierls-type distortions.

The lower and upper Hubbard bands of Am-5$f$ orbitals are increased under compression, associated with the gradual enhancing 5$f$ occupancy. Thus the $f-c$ hybridization become stronger at higher pressure. To quantitatively depict the hybridization process, we evaluated the impurity hybridization function $\tilde{\Delta}(\omega)$ for the Am-5$f$ orbitals as follows:
\begin{equation}
\label{eq:rehybd}
\tilde{\Delta}(\omega)=-\frac{\texttt{Im} \Delta(\omega)}{\pi}.
\end{equation}

Figure~\ref{fig:hyb} presents the impurity hybridization function for both the $5f_{5/2}$ and $5f_{7/2}$ states. The spectral weight of the hybridization function near the Fermi level increases monotonically in the sequence: Am-I < Am-II < Am-III < Am-IV. This monotonic enhancement provides direct evidence for the progressive increase in itinerancy of the 
5$f$ electrons across the high-pressure phases.

\subsection{Atomic eigenstate probabilities}

\begin{figure*}[ht]
\includegraphics[width=\textwidth]{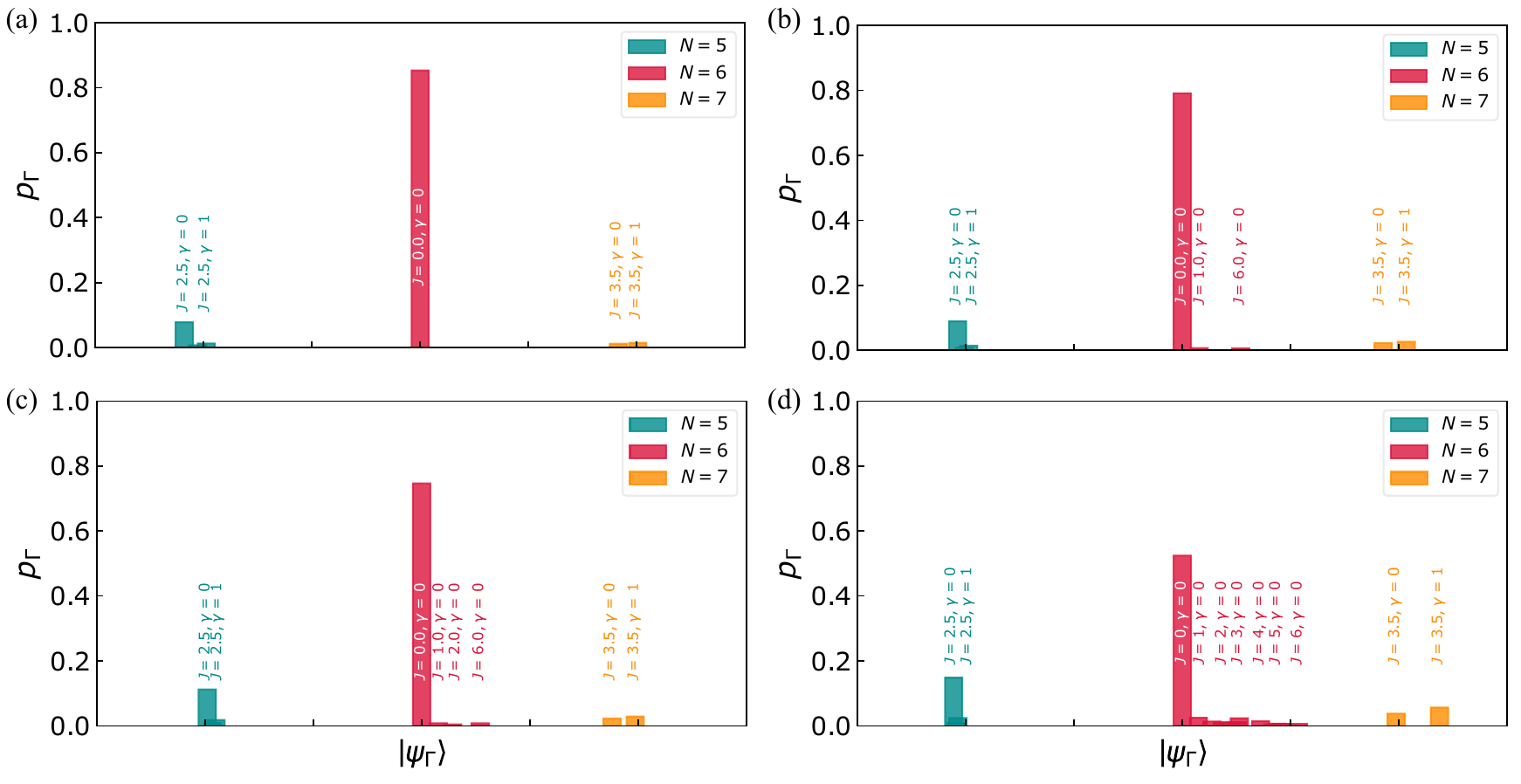}
\caption{(Color online) Valence state histograms of Am at $T = 290$~K by DFT + DMFT calculations. (a)–(d) Atomic eigenstate probabilities of Am-I, Am-II, Am-III, and Am-IV. The atomic eigenstates are denoted by using good quantum numbers $N$ (total occupancy), $J$ (total angular momentum) and $\gamma$ (rest of good quantum numbers), i.e., $|\psi_\Gamma\rangle=|N, J, \gamma \rangle$. The distributions of atomic eigenstate probabilities with respect to different $N$ are displayed in the legends. \label{fig:prob}}
\end{figure*}

The atomic eigenstate probability, or equivalently valence state histogram, serves as a powerful observable for examining the valence state fluctuation in strongly correlated materials~\cite{shim:2007}. It represents the probability $p_\Gamma$ of finding the valence electrons in a specific atomic eigenstates $|\psi_\Gamma \rangle$, which is typically labeled by good quantum numbers such as electron occupancy $N$ or total angular momentum $J$. A system exhibits weak or restricted valence state fluctuations if its valence electrons predominantly occupy only one or two atomic eigenstates (indicated by high corresponding probabilities)~\cite{PhysRevB.99.045109}. Conversely, if the electrons are distributed across a broad set of eigenstates with no single dominant configuration, valence state fluctuations are considered strong~\cite{Lawrence1981}.

To interpret the valence state fluctuations and mixed-valence behavior in Am, we extract the 5$f$ electron atomic eigenstates from the output of DMFT ground states. Here $p_\Gamma$ quantifies the probability for $5f$ electrons to reside in a given atomic eigenstate $\Gamma$. The average $5f$ occupancy is given by $\langle n_{5f} \rangle = \sum_\Gamma p_\Gamma n_\Gamma$, where $n_\Gamma$ is the electron count in each atomic eigenstate $\Gamma$. Finally, the probability of a specific $5f^n$ configuration is defined as $w(5f^{n}) = \sum_\Gamma p_\Gamma \delta (n-n_\Gamma)$.

Figures~\ref{fig:prob}(a)–(d) depict the calculated $5f^n$ configuration probabilities for Am, where only $n = 5, 6, 7$ contribute noticeably; other configurations have negligible probability. As summarized in Table~\ref{tab:n5f}, the probability of 5$f^6$ electronic configuration is overwhelmingly dominant, accounting for 86.9 \%, followed by the $5f^5$ and $5f^7$ electronic configurations in the Am-I phase. Under increasing pressure, the probability of the 5$f^6$ configuration gradually declines (Am-II: 82.7 \%, Am-III: 79.3 \%, Am-IV: 68.2 \%), while the probabilities of $5f^5$ and $5f^7$ configurations rise correspondingly (Am-IV: $w(5f^5) \sim 20.3$ \%, $w(5f^7) \sim 11.5$ \%). This shift signals enhanced valence state fluctuations in the Am-IV phase, with $5f$ electrons redistributing from the $5f^6$ configuration toward both $5f^5$ and $5f^7$ configurations, leading to a broad distribution across the three electronic configurations. These findings align with prior electrical resistivity measurements~\cite{EPL82.57007}.

To quantify the strength of valence state fluctuations, we introduce the parameter ${\cal V} = \sum_n P_n (P_n - 1)$, where $P_n$ denotes the occupancy number of 5$f$ electrons. Two limiting cases illustrate its meaning: ${\cal V}=0$ when $P_n=1$ for a single configuration, indicating the absence of valence state fluctuations and highly localized electrons; ${\cal V}=1$ corresponds to a perfectly dispersed distribution of configurations, representing maximal valence state fluctuations, a scenario rarely realized in real materials. The calculated values are ${\cal V}$ = 0.265, 0.365, 0.428, and 0.696 for Am-I, Am-II, Am-III, and Am-IV, respectively. Thus, the strength of valence state fluctuations increases in the sequence ${\cal V}(\text{Am-IV}) > {\cal V}(\text{Am-III}) > {\cal V}(\text{Am-II}) > {\cal V}(\text{Am-I})$.
This trend supports a consistent physical picture: in the low-symmetry, small-volume Am-IV phase under high pressure, $5f$ electrons show a clear tendency to participate in bonding, reflecting the onset of delocalization. Our results corroborate earlier studies~\cite{EPL82.57007,GRIVEAU200784} and help to clarify the controversy posed by X-ray absorption spectroscopy experiment, which had been interpreted as indicating localized $5f$ electrons in Am-IV~\cite{PhysRevB.82.201103}.

\subsection{Self-energy functions}

\begin{figure*}[ht]
\includegraphics[width=\textwidth]{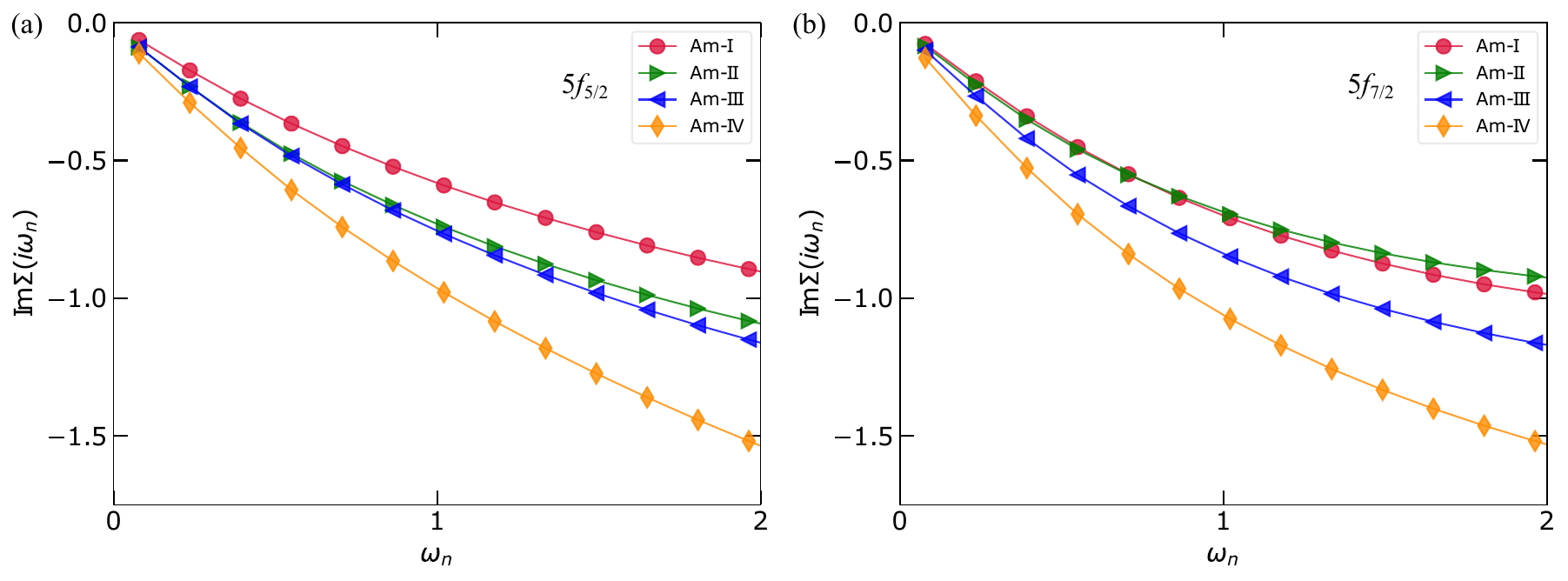}
\caption{(Color online) Imaginary parts of the Matsubara self-energy functions of Am at $T = 290$~K in the low-frequency regime by DFT + DMFT calculations. For the Am-5$f$ $j$=5/2 components. (b) For the Am-5$f$ $j$=7/2 components. \label{fig:sig}}
\end{figure*}

It is noteworthy that all electronic correlation effects are contained within the self-energy function. Figures~\ref{fig:sig}(a)–(b) illustrate the imaginary parts of the Matsubara self-energy functions at low frequencies for the 5$f$ orbitals across the four phases of Am. The concave shape of these low-frequency curves indicates metallic behavior. Furthermore, the low-energy electron scattering rate $\gamma$, given by the intercept on the $y$-axis, is smallest in Am-IV and largest in Am-I. Particularly, the scattering rates for the $5f_{7/2}$ states are consistently lower than those for the $5f_{5/2}$ states, although both manifolds follow a similar trend across the phases. Finally, it seems that the self-energy functions at low-frequency region exhibit a non-linear form, which deviates from the linear dependence predicted by Landau's Fermi-liquid theory~\cite{RevModPhys.68.13}.

The self-energy function $\Sigma(\omega)$ is utilized to determine both the quasiparticle weight $Z$ and effective electron masses $m^{*}$ for Am-$5f$ electrons~\cite{RevModPhys.68.13}:
\begin{equation}
\label{eq:renor}
Z^{-1} = \frac{m^{*}}{m_e} = 1 - \frac{\partial \texttt{Im}\Sigma(i \omega_0)}{\partial \omega_0},
\end{equation}
where $\omega_0 \equiv \pi/\beta$ and $m_e$ is the bare electron mass. The calculated orbital-resolved quasiparticle weights $Z$ and effective masses $m^{*}$ for the four phases of Am are summarized in Table~\ref{tab:sig}. The 5$f$ electrons in americium exhibit moderate correlation strength, with quasiparticle weights falling in the range $Z \in [0.4, 0.6]$ and effective masses ranging from $2.0m_e$ to $3.0m_e$. Remarkably, the $5f_{5/2}$ and $5f_{7/2}$ states in Am-I are more strongly renormalized, as reflected by their larger $Z$ values compared to the other three phases. To quantify the orbital differentiation between the $5f_{5/2}$ and $5f_{7/2}$ manifolds, we define the ratio $R \equiv Z_{5/2}/Z_{7/2}$. The calculated values yield $R \approx 1$ for Am-II, while $R \approx 0.9$ for the remaining phases, indicating only weak orbital-dependent correlations among the 5$f$ electrons in Am. This behavior contrasts with the moderately orbital-selective behavior observed in the 5$f$ electrons of Pu.

\begin{table}[th]
\caption{Calculated orbital-dependent quasiparticle weights $Z$ and electron effective masses $m^{*}$ for Am-I, Am-II, Am-III and Am-IV. \label{tab:sig}}
\begin{center}
\begin{tabular}{cccccc}
\hline
\hline
Phases & $Z_{5/2}$ & $Z_{7/2}$ & $m^{*}_{5/2}$ & $m^{*}_{5/2}$ & $R$ \\
\hline
Am-I   & 0.559 & 0.505 & 1.790$m_e$ & 1.980$m_e$ & 0.904 \\
Am-II  & 0.467 & 0.481 & 2.140$m_e$ & 2.080$m_e$ & 1.029 \\
Am-III & 0.472 & 0.440 & 2.120$m_e$ & 2.272$m_e$ & 0.933 \\
Am-IV  & 0.414 & 0.380 & 2.414$m_e$ & 2.630$m_e$ & 0.918 \\
\hline
\hline
\end{tabular}
\end{center}
\end{table}

\subsection{Angular momentum coupling scheme}

The evolution of 5$f$ electrons occupancy across the actinide series is a fundamental issue, as it dictates the angular momentum coupling of each element. In multielectron systems, the balance between spin-orbit coupling and Coulomb interaction gives rise to three distinct coupling regimes: Russell-Saunders (LS) coupling, $jj$ coupling, and intermediate coupling (IC)~\cite{RevModPhys.81.235}. While the ground states of late actinides are often described within an IC framework~\cite{PuBx2022}, the behavior for Am under high pressure remains an open and decisive question. Understanding how the 5$f$ occupancy and the associated angular momentum coupling evolve in the high-pressure phases of Am is essential, as it directly reflects the competition between localization, itinerancy, and relativistic effects under extreme compression.

The 5$f$ electronic states in actinides can be directly probed through core-level spectroscopic techniques, such as electron energy-loss spectroscopy (EELS) or X-ray absorption spectroscopy (XAS). Both methods involve exciting a core electron above the Fermi level, thereby probing the unoccupied density of states. When the excitation originates from a $d$ core level, the spin-orbit interaction per hole in the 5$f$ shell can be quantified using the spin–orbit sum rule. In such analyses, the branching ratio is extracted from the spin-orbit-split core edges, where dipole selection rules allow two types of transitions: $4d_{5/2} \rightarrow 5f$ and $4d_{3/2} \rightarrow 5f$ transitions. Because an excited $d$ core electron can only populate specific 5$f$ final states, the resulting differences in the branching ratio provide a measurable signature that can be rigorously interpreted within the spin-orbit sum-rule formalism. A key quantity derived from these measurements is the X-ray absorption branching ratio $\mathcal{B}$, defined as the relative intensity of the $4d_{5/2}$ absorption line~\cite{PhysRevA.38.1943}. This ratio serves as a direct experimental indicator of the spin-orbit coupling strength in the 5$f$ shell. Under the approximation that core-valence electrostatic interactions are neglected, $\mathcal{B}$ can be expressed analytically~\cite{shim:2007}. 

\begin{equation}
\label{eq:ratio}
\mathcal{B} = \frac{3}{5} - \frac{4}{15} \frac{1}{14 - n_{5/2} - n_{7/2}} \left ( \frac{3}{2} n_{7/2} - 2 n_{5/2} \right ),
\end{equation}
where $n_{7/2}$ and $n_{5/2}$ are the 5$f$ occupation numbers for the $5f_{7/2}$ and $5f_{5/2}$ states, respectively.
The calculated results are summarized in table~\ref{tab:n5f}. Compared with the previously reported values from K. T. Moore et al.~\cite{PhysRevB.76.073105}, where the occupancy of $n_{5/2}$=5.38 for $n_{7/2}$=0.62 yield an X-ray branching ratio $\mathcal{B}$=0.93. Our results give $n_{5/2}$=5.303 for $n_{7/2}$=0.625, corresponding to $\mathcal{B}$=0.919. The close agreement between the two sets of data confirms the reliability of our computational methodology and strongly supports the conclusion that the angular momentum coupling in Am approaches the $jj$ limit. It is noteworthy that while more delocalized 5$f$ states typically follow LS coupling, the strongly localized nature of the 5$f$ electrons in Am, together with the dominant occupancy of the $5f_{5/2}$ subshell, drives the system toward the $jj$ limit regime, resulting in a non-magnetic ground state with total angular momentum $J$=0.

\begin{table}[th]
\caption{The weights for 5$f$ electronic configurations ${w}(5f^n)$, 5$f$ orbital occupancy ($n_{5f}$, $n_{5/2}$ and $n_{7/2}$), total angular momentum $\langle J \rangle$ and X-ray absorption branching ratio $\mathcal{B}$ for Am-I, Am-II, Am-III and Am-IV. \label{tab:n5f}}
\begin{center}
\begin{tabular}{ccccccccc}
\hline
\hline
Phases & ${w}(5f^5)$ & ${w}(5f^6)$ & ${w}(5f^7)$ & $n_{5f}$ & $n_{5/2}$ & $n_{7/2}$ & $\langle J \rangle$ & $\mathcal{B}$ \\
\hline
Am-I   & 10.1\% & 86.9\% & 2.9\% & 5.927 & 5.303 & 0.625 & 0.424 & 0.919 \\
Am-II  & 11.7\% & 82.7\% & 5.5\% & 5.938 & 5.236 & 0.702 & 0.634 & 0.912 \\
Am-III & 14.9\% & 79.3\% & 5.8\% & 5.908 & 5.183 & 0.726 & 0.768 & 0.906 \\
Am-IV  & 20.3\% & 68.2\% & 11.5\% & 5.914 & 4.866 & 1.048 & 1.573 & 0.869 \\
\hline
\hline
\end{tabular}
\end{center}
\end{table}

\section{Discussion\label{sec:discussion}}

\textbf{High-pressure phase stabilized via Peierls-like distortion.}

\begin{figure*}[ht]
\includegraphics[width=\textwidth]{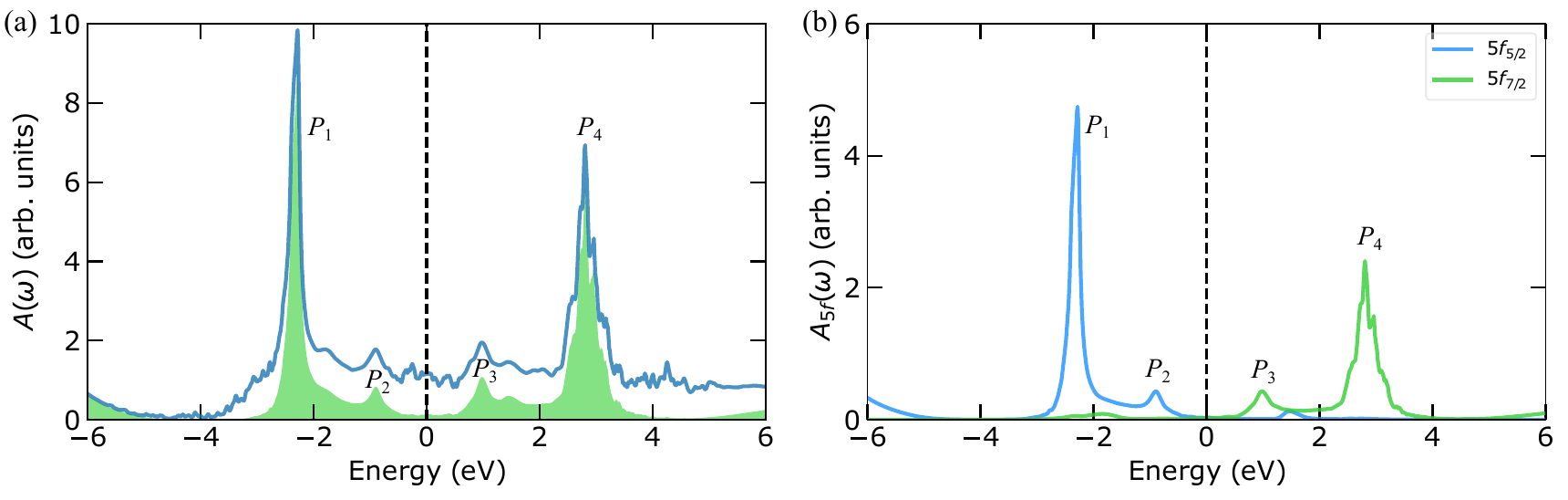}
\caption{(Color online) Total and 5$f$ partial density of states of Am-III at $T = 290$~K by DFT + DMFT calculations. (a) Total density of states A($\omega$) (in solid lines) and 5$f$ partial density of states A$_{5f}$($\omega$) (in colored shadow regions). (b) Orbital-resolved (or $j$-resolved) 5$f$ partial density of states $j$=5/2 and $j$=7/2. The Fermi levels $E_{F}$ are represented by vertical dashed lines. \label{fig:dosAmIII}}
\end{figure*}

The Peierls distortion plays a central role in stabilizing the low-symmetry high-pressure phases of americium, such as Am-III and Am-IV. As shown in Fig.~\ref{fig:dosAmIII}, the total and partial 5$f$ density of states of Am-III exhibits a distinct multi-peak structure near the Fermi level, consisting of four peaks labeled $P_1$–$P_4$. Specifically, $P_1$ and $P_4$ are centered at –2.3~eV and 2.8~eV, respectively, while $P_2$ and $P_3$ reside at –0.9~eV and 1.0~eV. These peaks arise from well-defined Am-5$f$ multiplet transitions. Although pressure enhances the itinerancy of 5$f$ electrons, their spectral weight near the Fermi level remains low, rendering the system a 5$f$ band insulator with narrow yet broadened bands. It is precisely under such conditions that a Peierls-like instability becomes favorable: the system lowers its total energy by breaking the crystal symmetry through a lattice distortion, which in turn splits the electronic bands and gives rise to the observed multi-peak density of states structure. This cooperative lattice-electron interaction not only generates the characteristic multi-peak density of states but also effectively reduces the electronic energy by occupying lower energy levels. The mechanism mirrors the charge-density-wave-driven Peierls distortion in $\alpha$-U~\cite{CDWUranium2016}, where symmetry reduction from $Cmcm$ to $Pbnm$ is likewise accompanied by electronic reconstruction and energy stabilization. Therefore, the high-pressure phase transitions in americium can be understood as a pressure-induced Peierls instability of the narrow 5$f$ bands, wherein symmetry-lowering lattice distortions and electronic reconfiguration act synergistically to minimize the total energy and stabilize the observed low-symmetry structures.

\section{Conclusions\label{sec:summary}}

In summary, this work provides a systematic and quantitative description of the correlated 5$f$ electronic states and phase stability of Am under pressure, spanning from the ambient-pressure Am-I to the high-pressure Am-II, Am-III, and Am-IV phases, using a charge fully self-consistent DFT + DMFT approach. Our calculations clarify a long-standing discrepancy between earlier transport experiments and X‑ray absorption studies by revealing a well-defined pressure-driven evolution of 5$f$ behavior: while the electrons remain strongly localized in Am-I and Am-II, they undergo progressive delocalization in Am‑III, marked by enhanced hybridization, transfer of spectral weight toward the Fermi level, and the emergence of multi-peak structures. In Am‑IV, this delocalization trend manifests as valence state fluctuations, with the 5$f$ occupancy redistributing from the dominant $5f^6$ configuration toward both $5f^5$ and $5f^7$ states. Furthermore, the evolution of the X‑ray branching ratio and angular momentum coupling across the phases is consistently explained. Crucially, we identify that the stability of the low-symmetry high-pressure phases is governed by a pressure-induced Peierls-like distortion of the narrow 5$f$ bands, which leads to characteristic multi-peak splitting in the electronic density of states. By offering a unified physical picture of 5$f$ electron localization, hybridization, and valence state fluctuations under compression, this study not only resolves the experimental-theoretical controversies but also establishes a coherent microscopic framework for understanding electronic and structural stability in heavy actinide systems under extreme conditions.

\begin{acknowledgments}
This work is supported by the National Key Research and Development Program of China (under Grant No.~2024YFA1408600), the National Natural Science Foundation of China (under Grant No.~12474241), Presidential Foundation of CAEP (under Grant No.~YZJJZQ2024014), National Key Research and Development Program of China (under Grant No.~2022YFA1402201).
\end{acknowledgments}

\bibliography{transU}

\end{document}